\documentclass[aps,showpacs,twocolumn,superscriptaddress]{revtex4-1}
\usepackage{amsmath}
\usepackage{amssymb}
\usepackage{graphicx}
\usepackage{epstopdf}
\usepackage{mathtools}
\newcommand{\ket}[1]{|#1\rangle}
\newcommand{\bra}[1]{\langle #1|}
\usepackage[bookmarks=false]{hyperref}
\hypersetup{colorlinks=true,citecolor=blue,linkcolor=blue,urlcolor=blue,pdfstartview=FitH,bookmarksopen=true}
\usepackage{xcolor}
\usepackage{soul}
\usepackage{times,txfonts}
\setstcolor{blue}
\setulcolor{red}

\begin{document}

\title{Experimental realization of nonadiabatic geometric gates with a superconducting Xmon qubit}
\author{P. Z. Zhao}
\affiliation{Department of Physics, Shandong University, Jinan 250100, China}
\author{Zhangjingzi Dong}
\affiliation{Department of Physics , Zhejiang University, Hangzhou 310027, China}
\author{Zhenxing Zhang}
\affiliation{Department of Physics , Zhejiang University, Hangzhou 310027, China}
\author{Guoping Guo}
\affiliation{Key Laboratory of Quantum Information, University of Science and Technology of China, Hefei, 230026, China}
\affiliation{Origin Quantum Computing, Hefei, 230026, China}
\author{D. M. Tong}
\email{tdm@sdu.edu.cn}
\affiliation{Department of Physics, Shandong University, Jinan 250100, China}
\author{Yi Yin}
\email{yiyin@zju.edu.cn}
\affiliation{Department of Physics , Zhejiang University, Hangzhou 310027, China}
\date{\today}

\begin{abstract}
Geometric phases are only dependent on evolution paths but independent of evolution
details so that they own some intrinsic noise-resilience features.
Based on different geometric phases, various quantum gates have been proposed, such as
nonadiabatic geometric gates based on nonadiabatic Abelian geometric phases and
nonadiabatic holonomic gates based on nonadiabatic non-Abelian geometric phases.
Up to now, nonadiabatic holonomic one-qubit gates have been experimentally demonstrated
with the supercondunting transmon, where three lowest levels with cascaded configuration are
all applied in the operation. However, the second excited states of transmons have
relatively short coherence time, which results in a lessened fidelity of quantum gates.
Here, we experimentally realize Abelian-geometric-phase-based nonadiabatic geometric
one-qubit gates with a superconducting Xmon qubit.
The realization is performed on two lowest levels of an Xmon qubit and thus avoids
the influence from the short coherence time of the second excited state.
The experimental result indicates that the average fidelities of single-qubit gates
can be up to $99.6\%$ and $99.7\%$ characterized by quantum process tomography and
randomized benchmarking, respectively.
\end{abstract}

\maketitle

\section{Introduction}

The implementation of circuit-based quantum computation requires to realize a universal
set of accurately controllable quantum gates, including arbitrary one-qubit gates and
a nontrivial two-qubit gate \cite{Bremner}.
However, the errors resulting from the imperfect control of quantum systems inevitably affect quantum gates.
This motivates researchers to utilize the features of geometric phases to suppress control errors.
Geometric phases are only dependent on evolution paths but independent of evolution
details so that the quantum gates based on geometric phases own some intrinsic noise-resilience
features \cite{Albert2003,Carollo,Thomas,Zhu2005,Thomas2011,Johansson}.
The early schemes of geometric-phase-based quantum computation \cite{Zanardi,Jones,Duan} is
realized by using adiabatic geometric phases \cite{Berry,Wilczek}. However, the long run
time required by the adiabatic evolution makes quantum gates vulnerable to the environment-induced decoherence.
To overcome this problem, nonadiabatic geometric quantum computation \cite{Wang,Zhu} based
on nonadiabatic Abelian geometric phases \cite{Aharonov} was proposed and
further nonadiabatic holonomic quantum computation \cite{Sjoqvist,Xu} based on nonadiabatic
non-Abelian geometric phases \cite{Anandan} was put forward.

Nonadiabatic Abelian geometric phases are generated by the cyclic evolution of quantum states
with removal of dynamical phases and the computational space itself is enough for the realization
of nonadiabatic geometric gates. Correspondingly, nonadiabatic non-Abelian geometric phases
are generated by the cyclic evolution and parallel transport of a state subspace in the Hilbert
space so that the realization of nonadiabatic holonomic gates needs the computation space
in addition to auxiliary states.
Since nonadiabatic geometric quantum computation as well as nonadiabatic holonomic quantum
computation not only has some intrinsic noise-resilience features but also allow high-speed
implementation, they have received increasing attention \cite{Zhu2003,Feng,Ota,Spiegelberg2013,Xu2014PRA,Liang2014,Xue2015,
Xu2015,Sjovist2016,Albert,Zhao2016,Sjovist2016PRA,Zhao,Zhao2017,Zhao2018,Zhao2019,Leibfried,Du,Long,Long2017,
Abdumalikov,Sun,ZhangzxNJP19,Danilin,Egger,Duan2014,Arroyo,Zhou2017,Sekiguchi,Nagata,Ishida}.
Up to now, nonadiabatic geometric gates have been experimentally demonstrated with trapped
ions \cite{Leibfried} and nuclear magnetic resonance \cite{Du}, and nonadiabatic holonomic gates
have been experimentally demonstrated with nuclear magnetic resonance \cite{Long,Long2017},
superconducting \cite{Abdumalikov,Sun,ZhangzxNJP19,Danilin,Egger}, and nitrogen-vacancy centers in
diamond \cite{Duan2014,Arroyo,Zhou2017,Sekiguchi,Nagata,Ishida}. Besides, adiabatic geometric
gates have been sped up \cite{Zhang2015,Liang2016} by using transitionless quantum driving
algorithm \cite{Berry2009} and the corresponding experiments have demonstrated this result \cite{Arroyo2018,Yan,WangthNJP18}.

The superconducting circuit provides an appealing experimental platform for the implementation
of quantum computation since the integrated circuit can be easily scaled up to multi-qubit
systems and each qubit can be controlled by individual lines. As a charge-insensitive
superconducting qubit \cite{Koch}, the transmon has been used for the experimental
realization of nonadiabatic holonomic one-qubit gates \cite{Abdumalikov,Sun,ZhangzxNJP19,Danilin},
where three lowest levels with cascaded configuration are used as a quantum system.
However, the second excited states of transmons have relatively short coherence time,
which results in a lessened fidelity of quantum gates. This motivates us to use nonadiabatic
Abelian geometric phase to experimentally realize nonadiabatic geometric gates since the
realization only needs a two-level computational space and does not require the extra auxiliary
level.

In this paper, we experimentally realize nonadiabatic geometric one-qubit gates with a
cross-shaped transmon qubit, named Xmon qubit \cite{Yin2013,Barends,Kelly}.
The superconducting Xmon qubit is a high-quality qubit because it not only possesses
relatively long coherence time but also balances coherence, connectivity, and fast
control with specific designing. Our realization is performed on two lowest levels of the
Xmon qubit and thus avoids the influence from the short coherence time of the second excited state.
By using the slice-shaped evolution path of two orthogonal quantum states \cite{Zhao2017}, we
can realize an arbitrary nonadiabatic geometric one-qubit gate by a single loop.
In our experiment, we demonstrate some specific nonadiabatic geometric gates.
The average fidelities of the gates can be up to $99.6\%$ and $99.7\%$ characterized by
quantum process tomography and randomized benchmarking, respectively.

\section{The protocol}

Our protocol is based on nonadiabatic Abelian geometric phases. We first explain how
the nonadiabatic Abelian geometric phases arise \cite{Aharonov}.
Consider a quantum state $\ket{\psi(t)}$ exposed to the Hamiltonian $H(t)$
satisfying Schr\"{o}dinger equation $i\ket{\dot{\psi}(t)}=H(t)\ket{\psi(t)}$.
If the state $\ket{\psi(t)}$ undergoes cyclic evolution, the initial state $\ket{\psi(0)}$ acquires a phase
\begin{align}
\gamma(\tau)=i\int^{\tau}_{0}\langle\nu(t)\ket{\dot{\nu}(t)} dt-\int^{\tau}_{0}\bra{\psi(t)}H(t)\ket{\psi(t)}dt,
\end{align}
where $\ket{\nu(t)}$ is an auxiliary state satisfying $\ket{\psi(0)}=\ket{\nu(0)}=\ket{\nu(\tau)}$.
Here, the first part of the phase $\gamma(t)$ is known as geometric phases and the second part
is known as dynamical phases. Further, if the dynamical phase is equal to zero, then $\gamma(\tau)$ yields a pure geometric phase.

We then demonstrate how to realize an arbitrary  single-qubit nonadiabatic geometric gate
with a superconducting Xmon qubit.
Consider a two-level Xmon qubit consisting of ground state $\ket{0}$ and the first excited state $\ket{1}$.
The transition $\ket{0}\leftrightarrow\ket{1}$ is facilitated by a resonant microwave pulse
with time-dependent Rabi frequency $\Omega(t)$.
In the rotating frame, the Hamiltonian under the rotating wave approximation reads
\begin{align}
H(t)=\Omega(t)\ket{0}\bra{1}+\mathrm{H.c.},
\end{align}
where $\mathrm{H.c.}$ represents the Hermitian conjugate terms.

\begin{figure}[t]
  \includegraphics[scale=0.5]{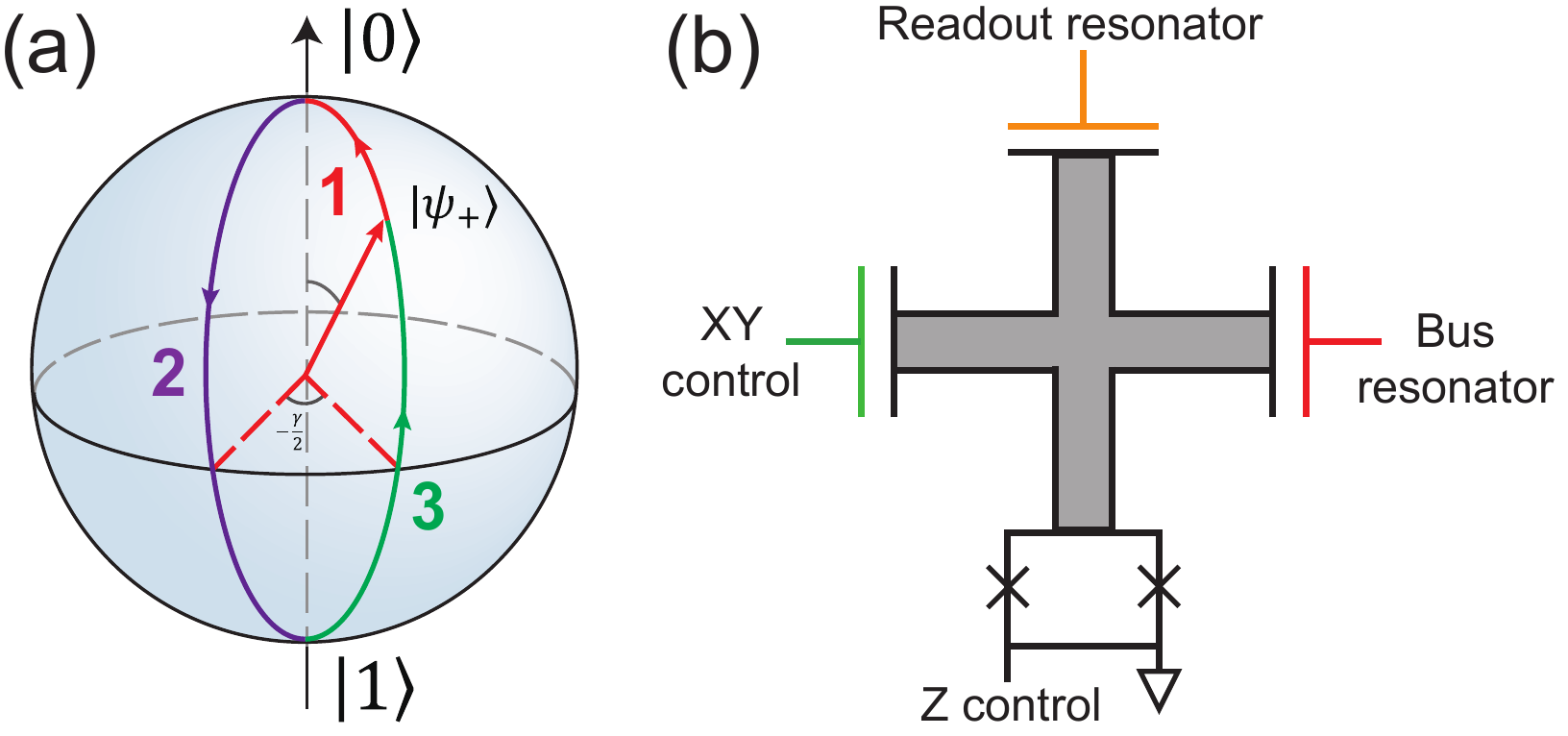}
  \caption{(Color online) The schematic setup for realization of nonadiabatic geometric gates.
  (a) The evolution path of state $|\psi_{+}\rangle$ with Bloch sphere representation. The
  state $|\psi_{+}\rangle$ is driven by the designed microwave pulses along an orange
  slice-shaped path in the computational space and acquires a geometric phase $-\gamma/2$.
  (b) The cross-shaped architecture of superconducting Xmon.
  }\label{Fig1}
\end{figure}

To realize the nonadiabatic geometric gates, we divide the whole evolution period $\tau$ into
three intervals $0\sim\tau_{1}$, $\tau_{1}\sim\tau_{2}$ and $\tau_{2}\sim\tau$, each of which
is with a different microwave phase parameter. The Hamiltonian in the first interval $0\sim\tau_{1}$
and the third interval $\tau_{2}\sim\tau$ is taken as
\begin{align}
H_{1}(t)=\Omega_{R}(t)e^{-i(\varphi-\frac{\pi}{2})}\ket{0}\bra{1}+\mathrm{H.c.},
\end{align}
and the Hamiltonian in the second interval $\tau_{1}\sim\tau_{2}$ is taken as
\begin{align}
H_{2}(t)=\Omega_{R}(t)e^{-i(\varphi-\frac{\gamma}{2}+\frac{\pi}{2})}\ket{0}\bra{1}+\mathrm{H.c.},
\end{align}
where $\Omega_{R}(t)$ is the time-dependent microwave amplitude parameter, and $\varphi$
and $\gamma$ are the time-independent microwave phase parameters.
Meanwhile, the evolution time $\tau_{1}$, $\tau_{2}$ and $\tau$ are taken to satisfy
\begin{align}
&\int^{\tau_{1}}_{0}\Omega_{R}(t)dt=\frac{\theta}{2},~\int^{\tau_{2}}_{\tau_{1}}\Omega_{R}(t)dt=\frac{\pi}{2}, \notag\\
&\int^{\tau}_{\tau_{2}}\Omega_{R}(t)dt=\frac{\pi}{2}-\frac{\theta}{2}.
\end{align}
With this design, two orthogonal states $\ket{\psi_{+}}=\cos(\theta/2)\ket{0}+\sin(\theta/2)\exp(i\varphi)\ket{1}$ and
$\ket{\psi_{-}}=\sin(\theta/2)\exp(-i\varphi)\ket{0}-\cos(\theta/2)\ket{1}$ undergo cyclic evolution,
\begin{align}
&\ket{\psi_{+}}\rightarrow U(\tau)\ket{\psi_{+}}=e^{-i\frac{\gamma}{2}}\ket{\psi_{+}},
\notag\\
&\ket{\psi_{-}}\rightarrow U(\tau)\ket{\psi_{-}}=e^{i\frac{\gamma}{2}}\ket{\psi_{-}},
\label{eq1}
\end{align}
and they respectively acquire phases $-\gamma/2$ and $\gamma/2$, where $U(\tau)$ is the final evolution operator reading
\begin{align}
U(\tau)=e^{-i\int^{\tau}_{\tau_{2}}H_{1}(t)dt}e^{-i\int^{\tau_{2}}_{\tau_{1}}H_{2}(t)dt}e^{-i\int^{\tau_{1}}_{0}
H_{1}(t)dt}.
\label{eq2}
\end{align}
We can verify that the dynamical phases $\gamma_{d_\pm}$ acquired by $\ket{\psi_{\pm}}$ are equal to
zero, i.e., $\gamma_{d_{\pm}}=-\int^{\tau_{1}}_{0}\bra{\psi_{\pm}(t)}H_{1}(t)\ket{\psi_{\pm}(t)}dt
-\int^{\tau_{2}}_{\tau_{1}}\bra{\psi_{\pm}(t)}H_{2}(t)\ket{\psi_{\pm}(t)}dt
-\int^{\tau}_{\tau_{2}}\bra{\psi_{\pm}(t)}H_{1}(t)\ket{\psi_{\pm}(t)}dt=0$,
where $\ket{\psi_{\pm}(t)}=U(t)\ket{\psi_{\pm}}$ with $U(t)$ being the evolution operator.
Therefore, $-\gamma/2$ and $\gamma/2$ are pure geometric phases.
By using the visualization Bloch sphere representation for the evolution of state
$|\psi_{+}\rangle$, $\gamma$ is indeed proportional to the solid angle enclosed by the
orange-slice-shaped loop, shown in Fig. \ref{Fig1}(a).
From Eq. (\ref{eq1}), we can readily find that the evolution operator $U(\tau)$ can be equivalently expressed as
\begin{align}
U(\tau)=e^{-\frac{\gamma}{2}}\ket{\psi_{+}}\bra{\psi_{+}}
+e^{\frac{\gamma}{2}}\ket{\psi_{-}}\bra{\psi_{-}}
=e^{-i\gamma\boldsymbol{n\cdot\sigma}/2},
\end{align}
where $\boldsymbol{n}=(\sin\theta\cos\varphi,\sin\theta\sin\varphi,\cos\theta)$ and
$\boldsymbol{\sigma}=(\sigma_{x},\sigma_{y},\sigma_{z})$ being the Pauli operators acting on $\ket{0}$ and $\ket{1}$.
As discussed above, $-\gamma/2$ and $\gamma/2$ are pure geometric phases, thus $U(\tau)$ is a nonadiabatic geometric gate.
By using this geometric gate $U(\tau)$, we can realize an arbitrary one-qubit gate, where the
rotation axis is determined by $\{\ket{\psi_{+}},\ket{\psi_{-}}\}$ and the rotation angle is determined by $\gamma$.

\section{The experiment}

In our experiment, the superconducting Xmon is an aluminum-based circuit, which is operated at temperature
about $10\mathrm{mK}$ in a cryogen-free dilution refrigerator. The Xmon sample is fabricated on a silicon
substrate using a standard nano-fabrication method. The Xmon device is designed with cross-shaped architecture,
shown in Fig. \ref{Fig1}(b). The four arms of Xmon device  connect to four separate elements with different
functions. The key element of the qubit is the pair of Josephson junctions at the bottom, which forme a rectangular
ring-shaped superconducting quantum interference device. The qubit frequency is tuned by the $Z$ control line.
The readout resonator on the top is for the measurement of qubit states.
The $XY$ control on the left is used for the manipulation of qubit states by inputting microwave pulses.
Through the right arm, the qubit can be coupled to a quantum bus resonator or another qubit.
Two lowest levels $\ket{0}$ and $\ket{1}$ of the Xmon are applied as the qubit states. The frequency difference of
qubit levels is $\omega_{10}/(2\pi)=5.266 \mathrm{GHz}$. The relaxation time is $T_{1}=19.0 \mathrm{\mu s}$
and the pure dephasing time is $T_{2}^*=10.0 \mathrm{\mu s}$. The readout fidelity of the ground state $\ket{0}$ and
excited state $\ket{1}$ are $F_q^0=98.0\%$ and $F_q^1=93.6\%$, respectively.

Our protocol allows to experimentally realize an arbitrary nonadiabatic geometric one-qubit gate by a single loop.
Our experiment specifically implements some certain nonadiabatic geometric gates, including identity
operator $I$, Hadamard gate $H$, and rotation gates $R_{x}(\pi)$, $R_{x}(\pi/2)$, $R_{y}(\pi)$, $R_{y}(\pi/2)$,
$R_{z}(\pi)$, as well as $R_{z}(\pi/2)$ with rotation axes denoted by $x,y,z$, and rotation angle denoted
by $\pi,\pi/2$. In the process of realization, we input a resonant microwave pulse to manipulate the qubit.
The rotation axis of geometric gates is adjusted by tuning microwave amplitude parameter $\Omega_{R}(t)$,
microwave phase parameter $\varphi$, and evolution time $\tau_{1}$, $\tau_{2}$ and $\tau$. The rotation
angle is adjusted by tuning microwave phase parameter $\gamma$.
For each interval, $\Omega_{R}(t)$ is designed as $\Omega_{R}(t)=\Omega_{0}\sin^{2}(\pi t/T)$ with the evolution
period being taken as $T=10\mathrm{ns}$. With this specific design of $\Omega_{R}(t)$, the microwave pulse
can be smoothly turned on and off at $\Omega_{R}(0)$ and $\Omega_{R}(T)$.

To characterize the performance of these nonadiabatic geometric gates, we apply quantum process tomography.
\begin{figure}[t]
  \includegraphics[scale=0.5]{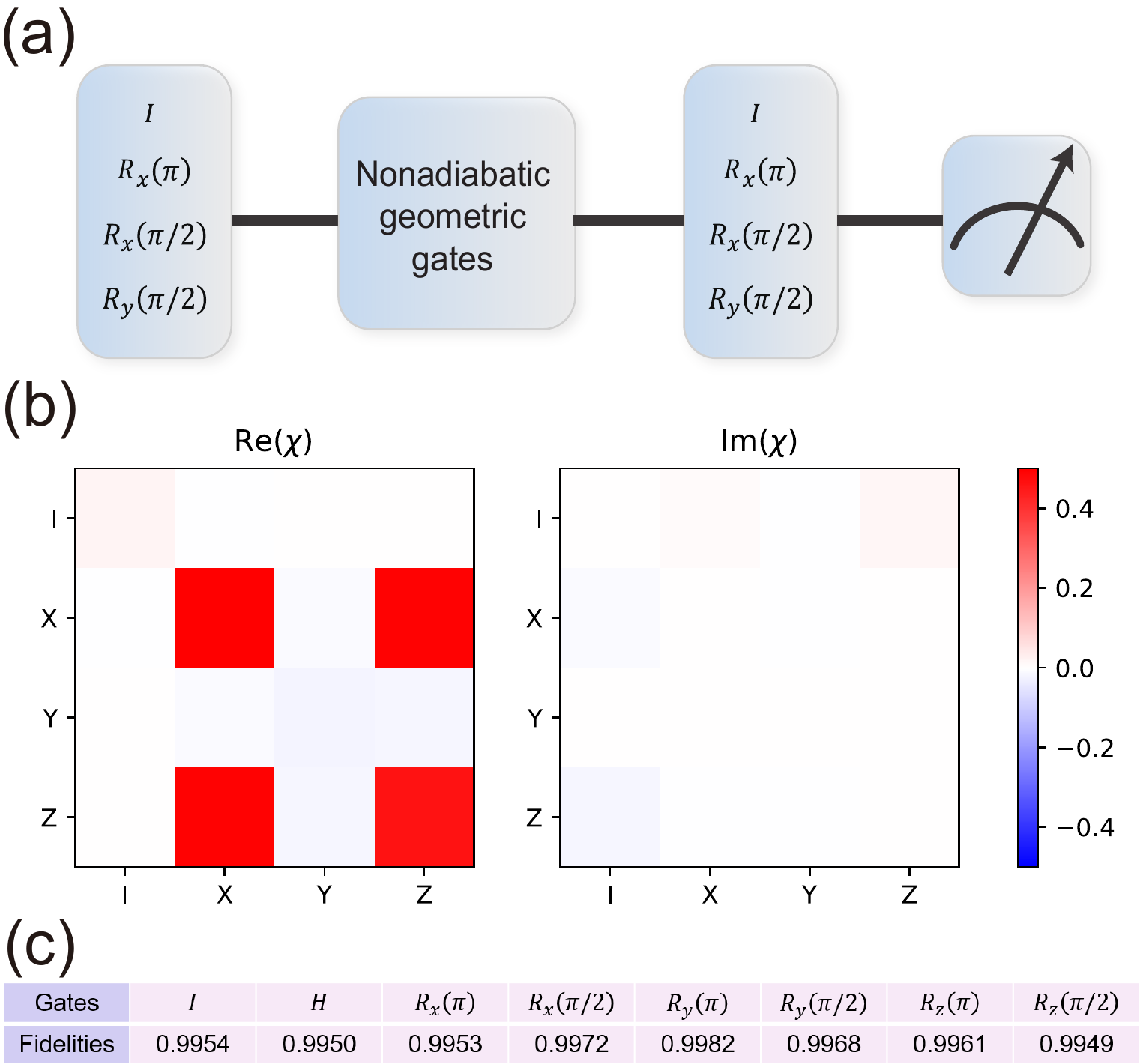}
  \caption{(Color online) Quantum process tomography for nonadiabatic geometric gates.
  (a) The schematic diagram for quantum process tomography.
  (b) The matrix elements generated by Hadamard gate $H$ with real part on the left and imaginary part on the right.
  (c) The process fidelities of nonadiabatic geometric gates.}\label{Fig2}
\end{figure}
In the quantum process tomography, we first prepare a set of initial states
$\{\ket{0},\ket{1},(\ket{0}-i\ket{1})/2,(\ket{0}+\ket{1})/2\}$, by applying a set of operators
$\{I,R_{x}(\pi),R_{x}(\pi/2),R_{y}(\pi/2)\}$ to the initialized state $\ket{0}$.
These states are then followed by a specific nonadiabatic geometric gate.
Finally, we perform measurement on the output states with standard quantum state tomography
and then reconstruct the output states.
The whole process is shown in Fig. \ref{Fig2}(a).
In this process, a set of fixed basis operators $E_{m}\in\{I,\sigma_{x},\sigma_{y},\sigma_{z}\}$
is chosen to describe the quantum dynamical map $\varepsilon(\rho_{i})=\sum_{mn}E_{m}\rho_{i}E_{n}\chi_{mn}$,
where $\chi_{mn}$ is the process matrix that completely characterizes the behavior of quantum dynamics.
By using the chosen initial states $\rho_{i}$ and the reconstructed output states $\varepsilon(\rho_{i})$,
we can calculate the process matrix $\chi_{mn}$.
Compared this process matrix $\chi_{mn}$ with the ideal process matrix $\chi_{id}$,
we can obtain the process fidelity $F_{\mathrm{P}}=\mathrm{Tr}(\chi_{mn}\chi_{id})$.
The matrix elements generated by the Hadamard gate $H$ is shown in Fig. \ref{Fig2}(b),
and the process fidelity $F_{\mathrm{P}}$ of the realized nonadiabatic geometric gates is
shown in Fig. \ref{Fig2}(c). Accordingly, we can obtain the average process fidelity $\bar{F}_{\mathrm{P}}=99.6\%$.
Please note that the measurement fidelity has been corrected in this fidelity to exclude the error from
qubit state readout.

Clifford-based randomized benchmarking is different measurement, from which the gate fidelity can
be separately extracted \cite{Chow,Magesan,Magesan2012}. Specifically, we perform reference
randomized benchmarking and interleaved randomized benchmarking.
The former gives an average fidelity of the Clifford gates and the latter gives the fidelity of a specific gate.

\begin{figure}[t]
  \includegraphics[scale=0.45]{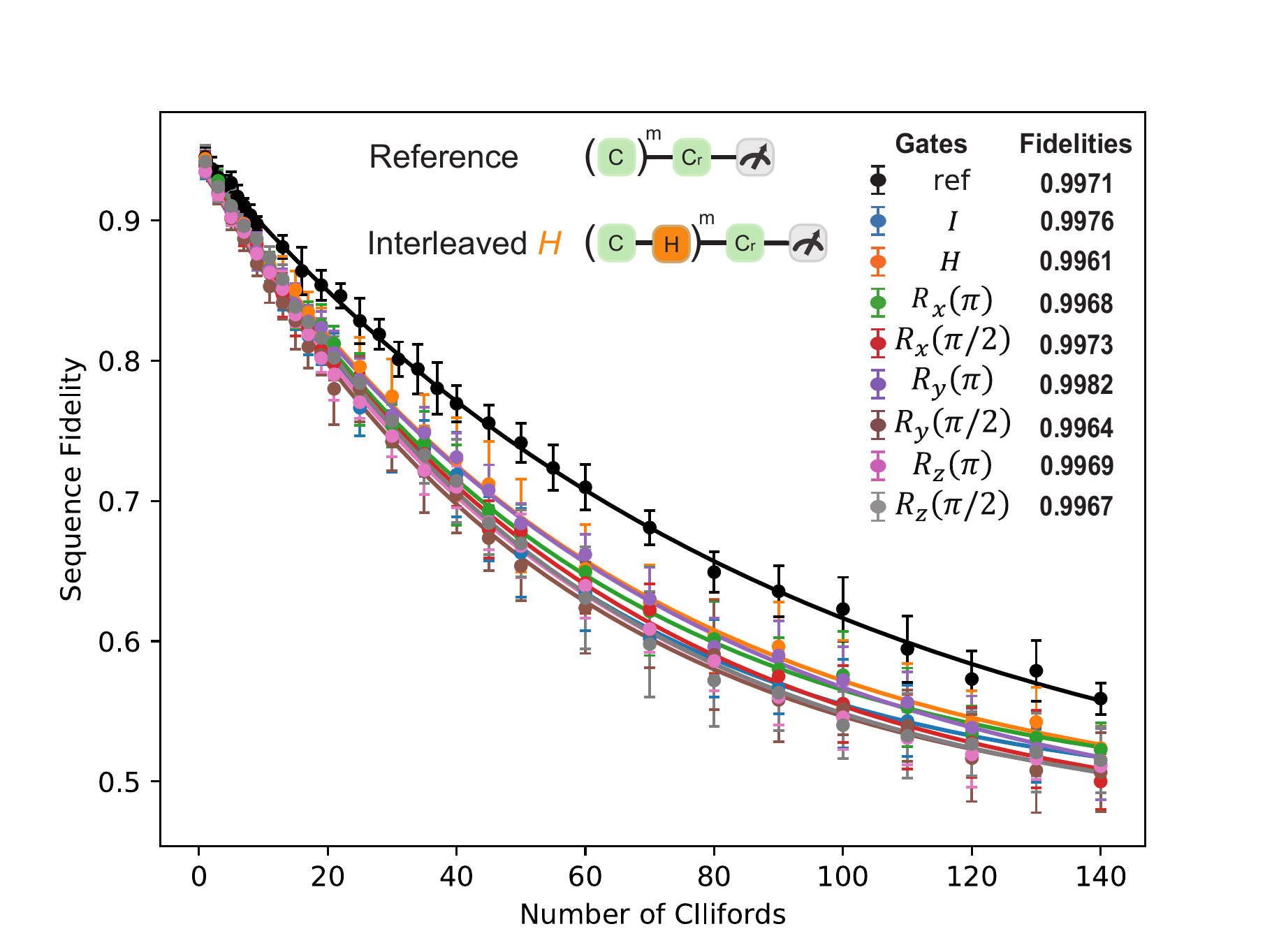}
    \caption{(Color online) Clifford-based randomized benchmarking of nonadiabatic geometric gates.}\label{Fig3}
\end{figure}

First, we perform the reference randomized benchmarking, the process of which is shown in Fig. \ref{Fig3}.
Specifically, we initially prepare quantum states in ground state $\ket{0}$, which is then driven by randomly chosen $m$ Clifford gates.
Subsequently, a recovery gate is applied to reverse the operation of the $m$ Clifford gates.
Afterwards, we measure the state $|0\rangle$ and obtain its survival probability.
Repeating this process by $50$ times, we can achieve the average survival probability of state $\ket{0}$ as the function of $m$, which is known as sequence fidelity.
The sequence fidelity $F$ can be fitted by using the function $F=Ap^{m}+B$, where $p$ is the depolarizing parameter, and $A$ and $B$ absorb the state preparation and measurement errors \cite{Magesan,Magesan2012}.
Accordingly, we can achieve the depolarizing parameter $p$.  The average error rate is then given by $r=(1-p)/2$. From the result of the reference randomized benchmarking experiment, shown in Fig. \ref{Fig3}, we obtain $p=0.994$, which yields average error rate $r=0.003$ and the average fidelity of nonadiabatic geometric Clifford gates $\bar{F}=99.7\%$.

Second, we perform the interleaved randomized benchmarking, the process of which is shown in Fig. \ref{Fig3}.
Specifically, we initially prepare quantum states in state $\ket{0}$. We then use $m$ combination of a randomly chosen nonadiabatic geometric Clifford gate and a target nonadiabatic geometric gate to drive this state.
Subsequently, a recovery gate is applied to reverse the above combination operation.  Afterwards, we measure the state $\ket{0}$ and obtain its survival probability.
With the similar procedure, we obtain a new depolarizing parameter $p_g$.
The fidelity of a specific gate can be then calculated by using the function $F_{g}=1-(1-p_{g}/p)/2$, where $p$ is the depolarizing parameter obtained in the  reference randomized benchmarking.
Figure \ref{Fig3} shows the fidelities of the realized nonadiabatic geometric gates.

Up to now, we have completed the quantum process tomography and the Clifford-based randomized benchmarking for a set of nonadiabatic gates, the average fidelity of which is $99.7\%$.

\section{Conclusion}

In conclusion, we have experimentally realized nonadiabatic geometric one-qubit gates $I$, $H$, $R_{x}(\pi)$, $R_{x}(\pi/2)$, $R_{y}(\pi)$, $R_{y}(\pi/2)$, $R_{z}(\pi)$, and $R_{z}(\pi/2)$ with superconducting Xmon qubits.  The average fidelities of these gates can be up to $99.6\%$ and $99.7\%$ characterized by quantum process tomography and randomized benchmarking, respectively. Different from nonadiabatic holonomic gates realized with three-level quantum systems, our experiment is performed on two lowest levels of an Xmon qubit and thus avoids the influence from the short coherence time of the second excited state. The methods and techniques used here can be extended to the other quantum systems, such as nuclear magnetic resonance, trapped ions, and nitrogen-vacancy centers in diamond.

\begin{acknowledgments}
The work reported here was supported by the National Basic Research Program of China (Grants No. 2015CB921004), the National Key Research and Development Program of China (Grant No. 2016YFA0301700), the National Natural Science
Foundation of China (Grants No. 11775129, No. 11575101, and No. 11934010), and the Anhui Initiative in Quantum Information Technologies (Grant No. AHY080000). This work was partially conducted at the University of Science and Technology of the China Center for Micro- and Nanoscale Research and Fabrication.
\end{acknowledgments}

\end{document}